\documentstyle[epsf,12pt]{article}
\begin{document}
\begin{titlepage}
\begin{center}
\vspace*{4cm}

\begin{title}
\bold {\Huge 
The Bose-Einstein effect in Monte Carlo  
      generators: weight methods}
\end{title}
\vspace{2cm}

\begin{author}
\Large {K. FIA{\L}KOWSKI\footnote{e-mail address:
uffialko@thrisc.if.uj.edu.pl},  R. WIT and J. WOSIEK }
\end{author}\\
\vspace{1cm}
{\sl Institute of Physics, Jagellonian University \\
30-059 Krak{\'o}w, ul.Reymonta 4, Poland}
\vspace{3cm}
\begin{abstract}
We present a method which  incorporates the Bose - Einstein 
effect into Monte Carlo
generators for multiple production by weighting the events. Various aspects of
weight calculations are discussed in detail. We show that our method allows to
describe reasonably well a sample of data and we outline the future tests and
applications.   
\end{abstract}
\end{center}

\vspace{1cm}

\noindent
TPJU-4/98  \\
March 1998 \\
hep-ph/9803399
\end{titlepage}

\section{Introduction}
\par
Recently there was much activity on the subject of the second order interference effects in
multiple production due to the Bose - Einstein statistics for pions [1]. In
particular, many authors have stressed the importance of incorporating these
effects into Monte Carlo generators which  well describe other features of multiple
production processes.
     It seems most natural to implement the Bose-Einstein (BE) effect  into Monte Carlo generators 
at the level of constructing the matrix element for the generation of
events. At present, however, this has been done only for a single Lund string [2] and for many 
models such an implementation seems to be  difficult. Therefore the standard 
procedure is to generate first the events according to the existing Monte Carlo
programs (without the BE effect), and 
then modify them. The most popular method [3] applies the momentum shifts for final 
state particles to reproduce the experimental two-body "Bose-Einstein ratio" 

\begin{equation}
\label{s0}
R_2(p_1, p_2) \  = \ {\rho _2(p_1, p_2 )  \over \rho _1(p_1)\rho _1(p_2)}.
\end{equation}

This method should be regarded as an imitation rather than an implementation of the effect,  
as it has no theoretical justification. The method has many other deficiencies, 
although some of them are recently claimed to be removed [4].
\par
   Another way of implementing the effect is to attach to the generated events
different weights, which 
depend on the momentum configuration, and thus to reconstruct the necessary 
enhancement in Bose-Einstein ratios. This method has a better theoretical 
justification. Following Pratt [5], Bialas and Krzywicki [6] have presented recently a 
derivation of the formula for weights based on the approximation for Wigner 
functions\footnote{A different method of implementing the BE
effect using Wigner functions (see  [7] and references quoted therein) requires a
Monte Carlo generator which defines for final state particles both momenta and
coordinates of the generation (or the last interaction) points in space.
Therefore it seems to be reliable only for heavy ion collissions. We will
not discuss it in this paper.}. This formula reads:

\begin{equation}
\label{s1}
W(n) = \sum_{\{P_n\}}\prod_{i=1}^{n}w_{i,k_i}.
\end{equation}                                                                         
\par
 Here $n$ is the number of identical particles, $k_i = P_n(i)$, $w_{i,k_i}$ is a two 
particle weight factor
calculated for the pair of momenta (of the $i-th$ particle and the particle which occupies
the $i-th$ place in the permutation ${P_n}$). The sum extends over all the
permutations of $n$ elements.
    In fact, such prescription should be applied separately for each sign of identical 
pions, and the full weight of the event is a product of three weight factors 
(one for each sign). 
The two-particle weight factor $w_{i,k}$ should be parametrized in a way reflecting the 
assumed space-time structure of the source. For pions coming from the decay 
of long-living resonances the effective source size is very large, and the effects 
appear for unmeasurably small differences of momenta. Thus only "direct" pions and 
the decay products of wide resonances for each event should be counted.
\par
The main difficulty with formula (2) is the factorial increase of the number of terms in 
the sum with increasing multiplicity of identical pions $n$ (other problems will be 
discussed later on). For high energies, when $n$ often exceeds 20, a straightforward 
application of formula (2) is impractical, and some authors [8,9] replaced it with 
simpler expresssions, motivated by some models. It is, however, rather difficult to 
estimate their reliability.
\par
     We have recently proposed two ways of dealing with this problem. One method 
consists of a truncation of the sum up to terms, for which the 
permutation $P(i)$ moves 
no more than 5 particles from their places [10]. This has a simple
 justification: for 
$P(i)=i$  a two-particle weight factor $w_{i,P(i)}$ is equal 1, and for non-equal indices it 
should decrease fast (usually Gaussian shape is assumed) with increasing 
difference between the momenta of  the $i$-th and the $P(i)$-th  particle. Thus the magnitude 
of terms, for which more than $k$ factors in the product are different from 1, should 
decrease quickly with $k$. We have checked for the $p{\overline p}$ minimum bias events at 630
GeV [10] and for the $e^+e^- \to W^+W^-$ events at LEPII [11] that 
changing the maximal value of k from 4 to 5 hardly influences the distributions, 
and in particular the two-particle ratio (1). However, it is difficult to claim a priori 
that such a truncation does not change the results which would be obtained using 
the full series (2).
\par
   Therefore a second way of an approximate calculation of the sum (2) was proposed 
[12]. Since this sum, called a permanent of a matrix built from weight factors
$w_{i,k}$, 
is quite familiar in field theory, one may use a known integral representation and 
approximate the integral by the saddle point method. This approximation improves in fact 
with increasing number of particles n. However, this method is reliable only if in each row 
(and column) of the matrix there is at least one non-diagonal element significantly 
different from zero (as already noted, all diagonal elements are equal to 1). 
Thus 
the prescription should not be applied to the full events, but to the
clusters, in which each 
momentum is not far from at least one other momentum. The full weight is then a 
product of weights calculated for clusters, in which the full event is divided. 
\par
   These considerations suggest the necessity of combining two methods. After dividing the 
final state momenta of identical particles into clusters, one should use for small 
clusters exact formulae presented in [10,11]. For large clusters (with more 
than five particles) one should compare two approximations (truncated series and 
the integral representation) to estimate their reliability and the sensitivity of
the final results to the method. Obviously, the results will depend also on the 
clustering algorithm used: if we restrict each cluster to particles very close in 
momentum space, the neglected contributions to the sum (2) from permutations 
exchanging pions from different clusters may be non-negligible, and if the cluster 
definition is very loose, the saddle point approximation may be unreliable. This should 
be then also checked to optimize the algorithm used.
\par
   In this paper we investigate and compare various approximations to formula 
(2) 
for the events generated by default PYTHIA/JETSET generator [13] for the  
proton-antiproton colisions at 630 GeV (simulating conditions of the  
UA1 experiment [14], for 
which many data on the BE effect were collected). In the next section we 
define used procedures and compare the weight distributions obtained with different 
assumptions. In Section 3 the problem of weight rescaling (necessary to reproduce 
the experimental multiplicity distributions) is discussed and some comparison with 
data is presented. We conclude with Section 4.

\section{Event generation and the procedures calculating weights }
\par 
   As already noted, we have chosen for our analysis the proton-antiproton collisions at 
CM energy of 630GeV.  We generate minimum bias events 
according to the default PYTHIA-JETSET generator [13]. The imitation  of the Bose-Einstein effect 
by momenta shifting is  switched off. 
For the trial runs, samples of 10 000 events were generated.
\par
 For each event the weight was 
calculated according to different procedures and parameter choices. In all cases we counted 
only the direct pions and 
the decay products of wide resonances (dominated by $\rho$). This is easily done if the procedures 
which calculate weights are called not after the generation of final state, but in the same place, in 
which the original LUBOEI procedure was called [3].
\par
The first procedure, referred  further to as "no-clustering with restricted
permutations" is the same as  applied 
before to the rough description of UA1 data [10] and to the discussion of W mass shifts
[11]. As already 
noted, it consists of separating in the sum (2) for each sign of pions into  the classes of  permutations 
which change places of exactly k momenta,

\begin{equation}
\label{s3}
W(n) = \sum_kW^{(k)}(n),
\end{equation}
and neglecting all terms with $k>k_{max}$. As previously, we used $k_{max}=5$. 
The detailed formulae used to calculate this approximation were given before [10,11]. The 
full weight of the event is the product of three weight factors calculated for 
pions of each  sign. 
\par
The shape of the two-particle weight factor $w_{i,j}$ in (2) should be chosen to fit the 
``BE ratio''.  For practical reasons we do not use the function $R_2(p_1,p_2)$
but the ratio of integrals  
\begin{equation}
c_2(Q) = \frac {\int d^3p_1d^3p_2 \rho _2(p_1,p_2)\delta [Q-
\sqrt{-(p_1-p_2)^2}]} {\int d^3p_1d^3p_2
\rho _1(p_1)\rho _1 (p_2)\delta [Q-
\sqrt{-(p_1-p_2)^2}]} \frac {<n>^2}{<n(n-1)>}.
\end{equation}
which is a  function of a single variable 
$Q = \sqrt {-(p_1 -p_2)^2}$.
Motivated by many experimental fits we parametrize $w_{i,j}$ as
\begin{equation}
\label{s2}
w_{i,j} = e^{(p_i - p_j)^2/2\sigma ^2}
\end{equation}
Of course,  different components of momentum difference squared may be
multiplied by different coefficients, and the shape may be modified. In this
note we do not discuss these possibilities. Therefore the only parameter is a
Gaussian half-width of the distribution $\sigma$.                                      
As noted in [10] the values of $\sigma$ of 0.1-0.2 GeV reproduce quite well the experimental width 
of the Bose-Einstein enhancement, and we will use the values from this range.
\par
In two other procedures we separate first the momenta of pions of each sign into clusters. The 
pion is assigned to a given cluster, if  the (minus) square of the difference of its four-momentum with at 
least one pion from this cluster is smaller than the assumed value of a parameter
$\epsilon$. Then, for 
each cluster, a cluster weight factor is calculated by the formula (2) using one of two approximations 
to be described below, and the final weight factor for each sign is a product of all cluster weight 
factors. Obviously, if there is only one pion in the cluster, its weight factor equals one. Note that 
the results may depend strongly on the ratio $\epsilon / \sigma^2$ . If this ratio is too small, 
the neglected terms in 
(2) (corresponding to the permutations, which exchange pions from different clusters) may be 
quite big. If it is large, the clusters will contain on the average many particles and the computing time 
will increase. Moreover, the matrix built out of two-particle weight factors 
for a cluster may have zero modes, and the saddle point
approximation [12] becomes unreliable. Thus the results for different values of 
$\epsilon$ should be compared.
\par
If the number of pions in a cluster is smaller than six, the sum (2) is calculated exactly
with the formulae from refs. [10,11]. For (rather few) larger clusters two 
alternative approaches are used. In the first version, referred to as "clusters with restricted permutations", 
the same 
approximation to the sum (2) as described above is applied. Note that now, for the events 
containing many clusters with more than one pion, this takes into account permutations, which may
change places of many more  than five momenta. However, the global number of used permutations is usually 
lower than for this approximation applied to the full final state. In the second version, referred to as 
"clusters with the saddle point method" the method from ref. [12] is used.
\par
We have started with the parameter values of $\epsilon=0.1 \ GeV^2$ and $0.15 \ GeV^2$, and
$\sigma = 0.17 \ GeV$. The first 
observation is that the original "no clustering" algorithm needs much more computing 
time (here by an order of magnitude) than the two others. Moreover both algorithms with restricted 
permutations
approximate the full weight factor for each sign  from below, since in both cases positive 
terms in the full sum (2) are neglected. Thus it is evident that the approximation which gives bigger weights is 
better. We find the average value (taken only for weight factors below 100) of 1.84 without 
clustering, and 1.88 or 1.90 for "clusters with restricted permutations"
 with the two values of $\epsilon$ quoted above. Even though these differences seem small, we feel that with 
 our statistitics they are significant. Moreover, without clustering
 there are only five cases (out of 30 000) of weight factors above 100, and with clustering there are 
13 or 14 such cases. We see that the procedure with clustering gives slightly larger values of weights, which means 
that it approximates the full sum better than the other one, even though it neglects more terms and uses 
much less computer time. In the following we will thus discuss only procedures with clustering.
\par
   The comparison of weight factors calculated with these two procedures gives less unambiguous 
results. The average values of weight factors are bigger for "restricted permutations", but the number of cases of 
values above 100 is bigger for "saddle point method". More precisely, the average weight factor for the saddle
point method is 
1.82 or 1.83, and the number of values above 100 is 19 or 21 (to be compared with numbers 
quoted above for "restricted permutations"). However, analyzing in more detail the events with anomalously 
high weight values, and adding the results for smaller (and more realistic) value of $\sigma=0.14
\ GeV$ 
we find that the saddle point method gives larger values of weights only in these cases, when they are anyway 
very high. For smaller $\sigma $ for both procedures only in 3 cases (out of 30 000) the weight factor 
exceeds 100, and the averages are 1.35 and 1.33 for the two procedures. 
Thus the saddle point algorithm gives 
slightly better approximation of weight factor only for anomalous cases, which 
are not very relevant for the analysis, as will be discussed later. 
\par
We conclude that the main improvement consists of introducing clustering into the
analysis. With this modification both algorithms, i.e. "restricted
permutations" and  "saddle point approximation", have similar performance.
Restricted permutations gives slightly better average while the saddle point method
is advantegeous at reproducing large weights. The situation may change for the processes 
with different density of particles in phase-space, or after improving the approximations used in 
the saddle point method. In the following we use the restricted permutation algorithm.  
\par
 Since all factors in the sum (2) are positive and $w_{i,i} =1$,
the resulting weight is not smaller than one (a contribution from identity
permutation). One may rescale the weights to keep, e.g., the average number of
particles fixed; we return to this point later. 
\par
As already noted, we are using the final weight of the event
given by a product of weight factors calculated separately for
positive, negative and neutral pions.  In fact, the  BE interference for neutral
particles is not observable (apart from the possible effects for direct photons
[15]): neutral pions decay before
detection, and for the resulting photons the effective source size is so big
that the BE effects must be negligible for momentum differences above a few eV.
However, the procedure should not change the observable correlations between the
numbers of charged and neutral pions. Therefore weights for all signs of pions
must be really taken into account.
\par

\section{Results and comparison with data}
\par
Before using our prescription for weights to calculate the BE ratio (4) let us
 note that weights do change not only this ratio (which for the default 
JETSET/PYTHIA generator is always close to one in clear disagreement with data) but
also many other distributions. Thus with the default values of free model 
parameters (fitted to
the data without weights) we  find inevitably some discrepancies with data
after introducing weights.
\par 
We want to make clear that this cannot be taken as a flaw of the
weight method. There is no measurable world ``without the BE effect'', and it 
makes no sense to ask, if this effect changes e.g. the multiplicity
 distributions. If any model is compared 
to the data without taking the BE effect into account, the fitted values of its
free parameters are simply not correct. They should be refitted with weights, 
and then the weights recalculated in an iterative procedure. This, 
however, may be a rather tedious task.
\par
Therefore we use, as in the previous papers [10,11] a simple rescaling method
 proposed by Jadach and Zalewski [8]. Instead of refitting the free parameters
of the MC generator, we rescale the BE weights (calculated according to the
procedure outlined above) with a simple factor $cV^n$, where 
$n$ is the global multiplicity of "direct" pions, and $c$ and $V$ are fit parameters. Their 
values are fitted to minimize the differences between the multiplicity
distribution obtained from MC without weights, and the one obtained with the
rescaled weights. As noted before [10], one may even avoid fitting if the 
distributions without- and with weights (non-rescaled) are well parametrized 
with the negative binomial distributions (NBD) with parameters $<n>,k$ and 
$<n'>,k'$, respectively. 
In our case, however, the distribution with weights becomes rather jagged for
large multiplicities due to the few events with anomalously high weight values.
Therefore the values of NBD parameters would be unreliable and direct fitting
of $c$ and $V$ is necessary. 
\par
The results, however, are fortunately rather
insensitive to the ``anomalous'' cases. Even if we remove them from the fitting
procedure and then include again for rescaling, the rescaled weight
distribution exhibits an "anomalous" tail greatly reduced. Out of  300 000 values of
weight factors only 5 exceed 100. Moreover, the resulting multiplicity
distribution with rescaled weights is quite smooth and approximates well the
original distribution. Specifically, with $\epsilon$ and $\sigma$ values as
quoted above, the average multiplicity and dispersion do not change more than
by a few percent. This is illustrated in Fig.1 for still smaller (but realistic)
value $\sigma = 0.1 \ GeV$ 
and $\epsilon= 0.1 \ GeV^2$.  It should be contrasted with
the $20\%$ increase of  the average multiplicity which occurs 
 when weights are introduced without rescaling (not shown). 
\vspace{0.5cm}

\epsfxsize=12cm

~~~~~~~~~~~~~\epsfbox{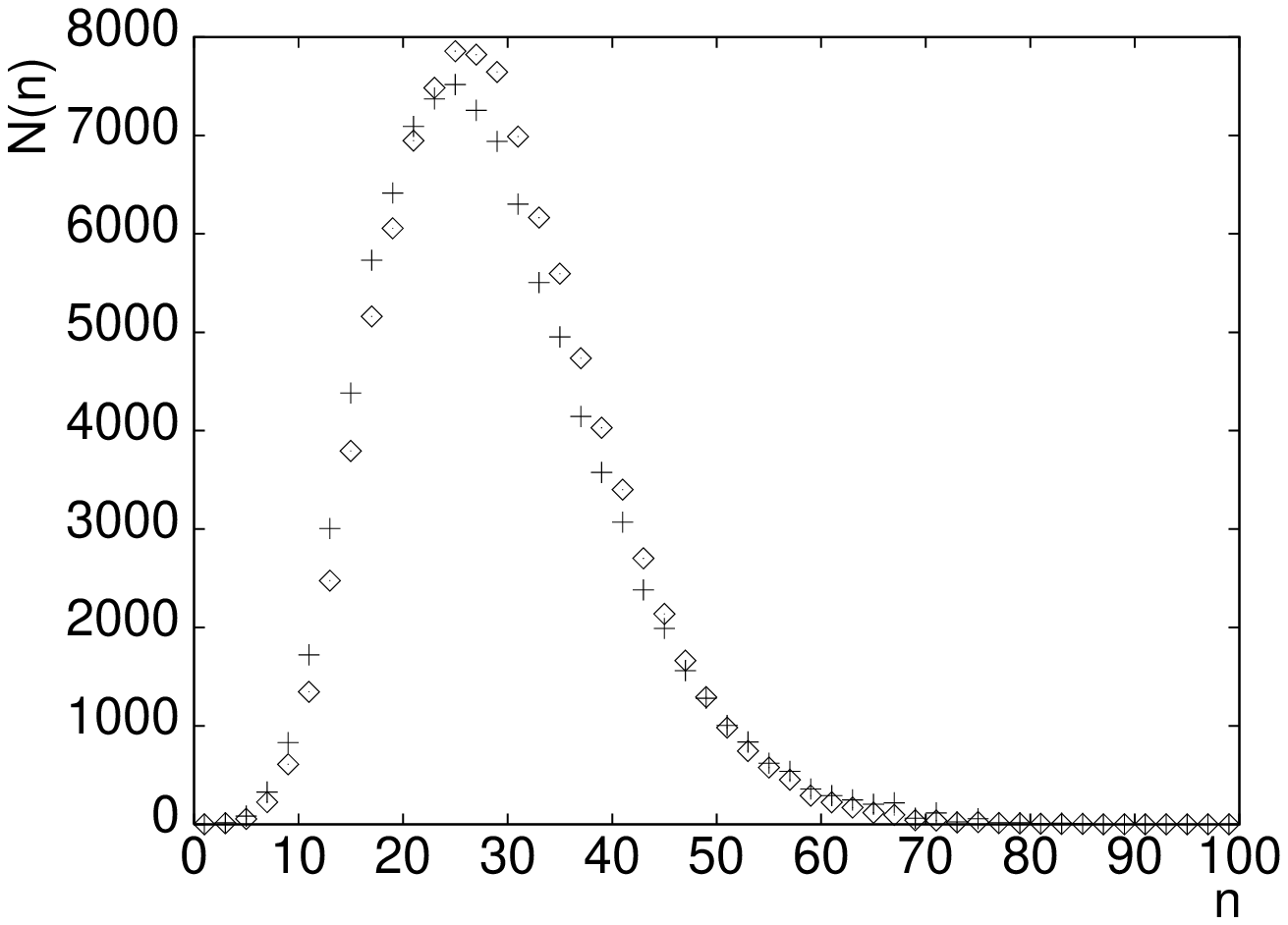}

\vspace{0.5cm}
\par
 {\bf Fig.1.} {\sl The  multiplicity distribution (in number of events)  without
weights (diamonds) and with rescaled weights (crosses).}  
\vspace{0.5cm}
\par
Moreover, the rescaling in $n$ gives also satisfactory corrections to the 
inclusive distributions in transverse- and longitudinal momenta (or Feynman $x$).
 Whereas the unrescaled weights shift the average value of $ln(1/x)$ by about 
0.7, rescaling restores the original value with the accuracy of a few percent. 
The corresponding distributions of $ln(1/x)$ are shown in Fig.2.
\vspace{0.5cm}
\epsfxsize=12cm

~~~~~~~~~~~~~\epsfbox{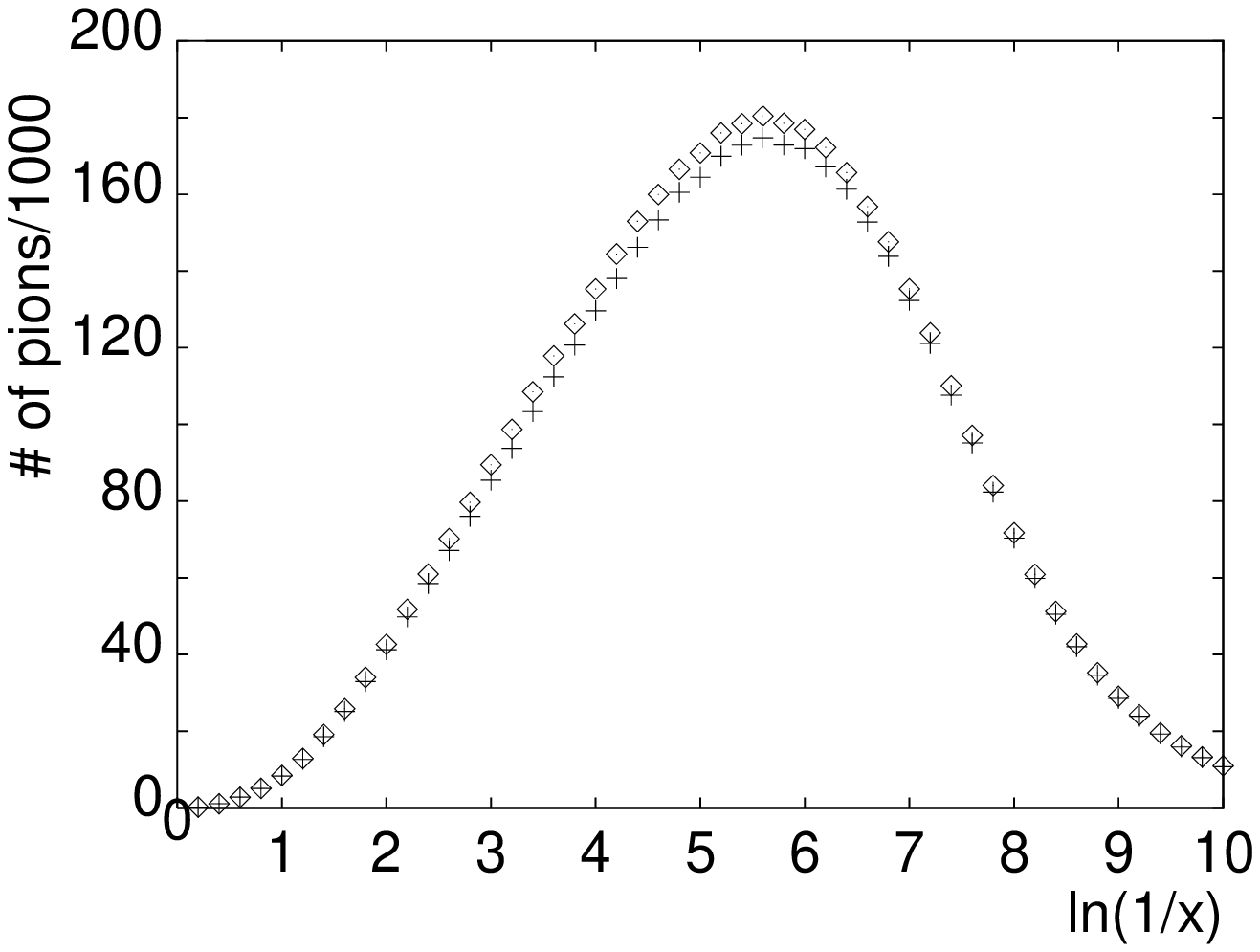}

\vspace{1cm}
\par
 {\bf Fig.2.} {\sl The distribution  of $ln(1/x)$ (in thousands of pions) without weights (diamonds)
and with rescaled weights (crosses).}   \\
\par
Similar effect is seen for transverse momenta, as shown in Fig.3.
 
\vspace{0.5cm}
\epsfxsize=12cm

~~~~~~~~~~~~~\epsfbox{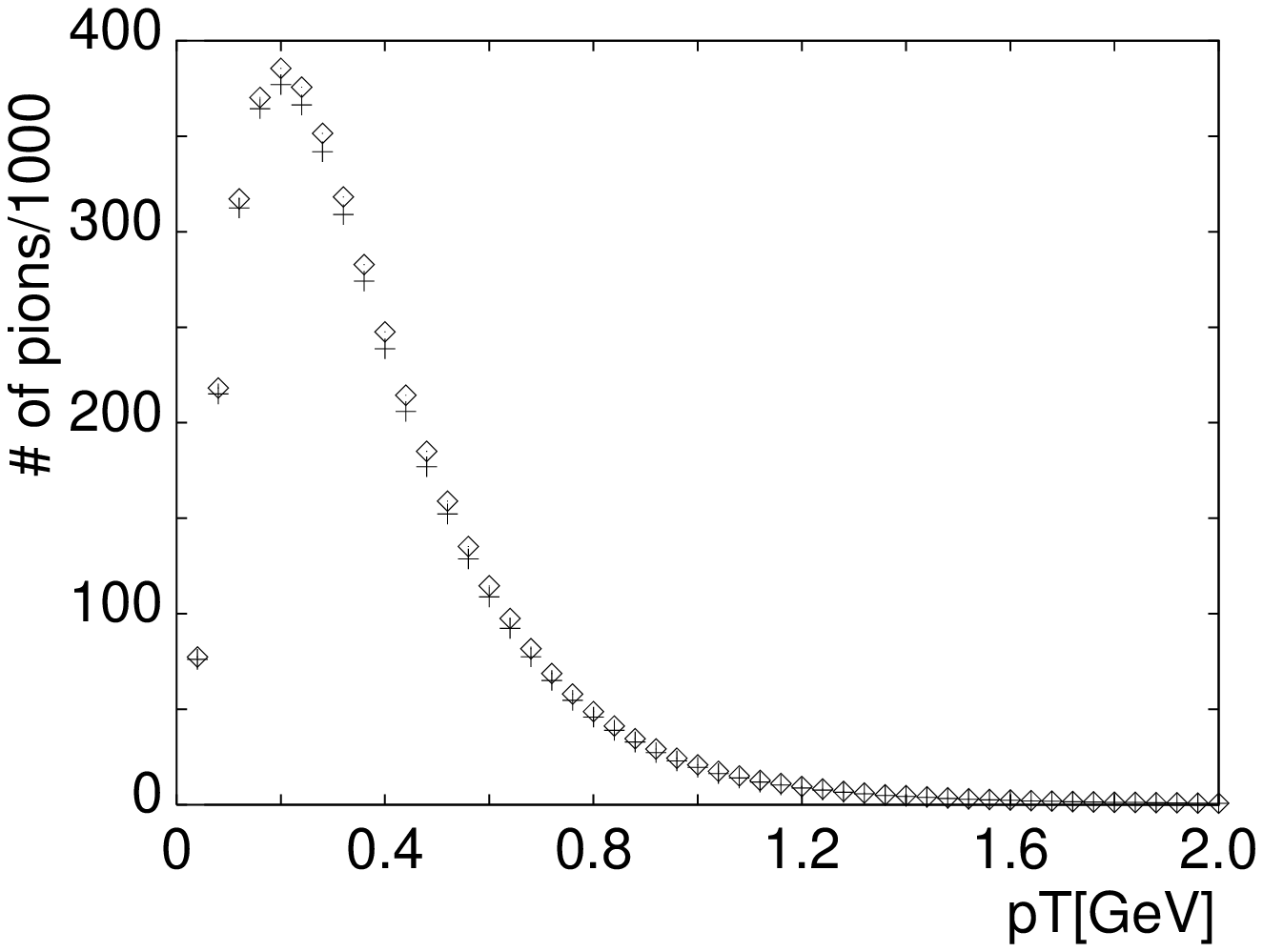}

\vspace{1cm}
\par
 {\bf Fig.3.} {\sl The transverse momentum distribution (in thousands of pions) without 
weights (diamonds) and with rescaled weights (crosses).}\\

 Thus it seems to be sufficient
to rescale weights with respect to a single quantity $n$; the single-particle
distributions recover then authomatically their original shape. Let us note
that, as we have already checked before [11], the shape of BE ratio (4) is
practically the same with- and without rescaling. This suggests that the
rescaling was indeed well chosen and corresponds to the refitting of such
parameters, which do not influence the BE ratio. 
\par
Obviously, for the  more detailed analysis of the final states,
single rescaling may be not enough. E.g., since different parameters govern
the average number of jets, and the average multiplicity of a single jet,
both should be rescaled separately to avoid discrepancy with the data. Let us
stress once again, however, that such problems arise only due to the use of
generators with improperly fitted free parameters, and do not suggest any
flaw of the weight method.
\par
Now we may check if the new version of our procedure yields the same Bose - Enstein
ratio (4) as the old one. In Fig.4 we show that within the statistical
fluctuations this is indeed the case.     
\vspace{0.5cm}

\epsfxsize=12cm

~~~~~~~~~~~~\epsfbox{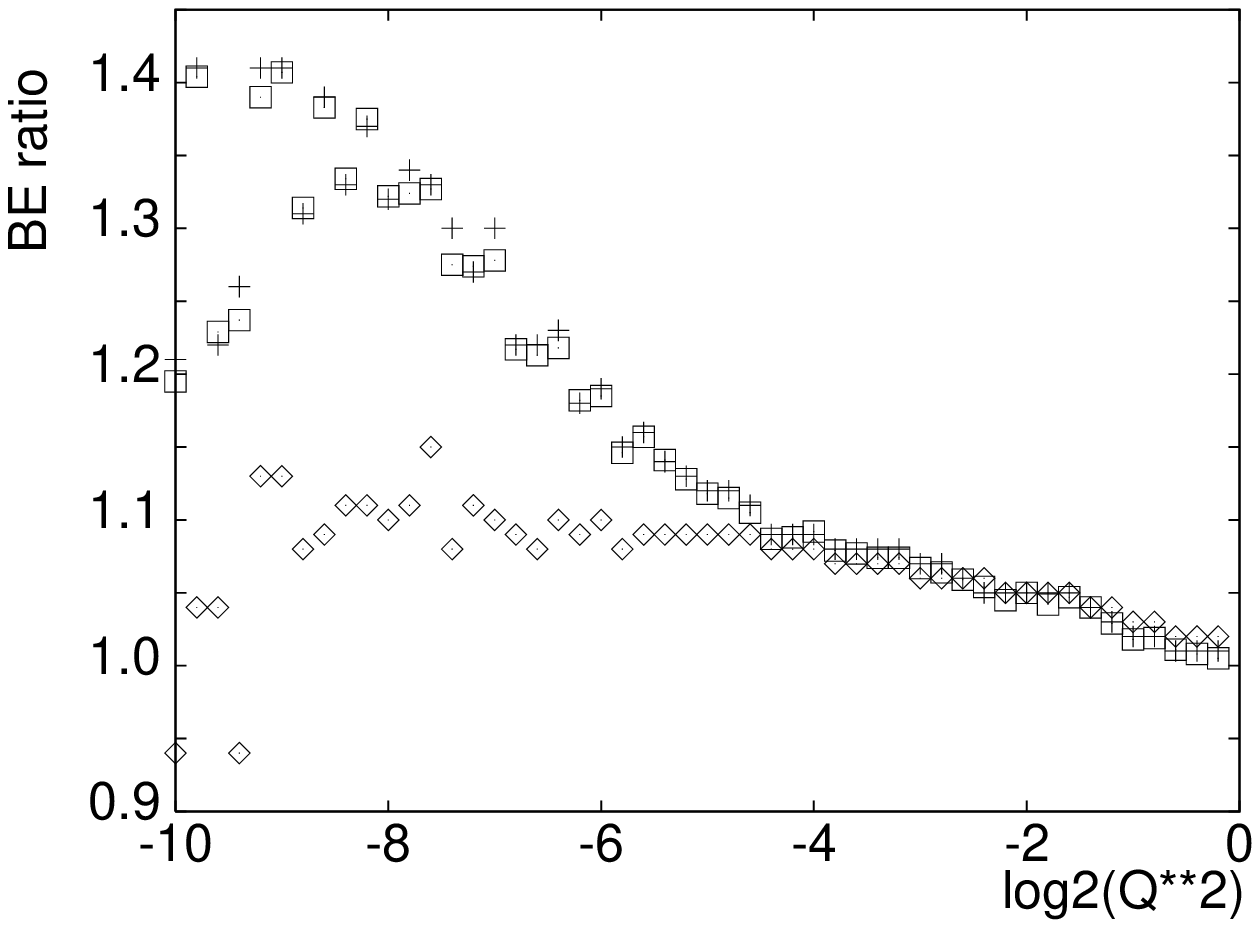}

\vspace{0.5cm}
\par
 {\bf Fig.4.} {\sl The  BE ratio (4) as a function of
 $log_2(Q^2/1GeV^2)$. 
Crosses, squares and diamonds correspond to the results with weights 
from the new procedure,
from the old one, and without weights, respectively. 
} \\

\par
 Since the data of UA1  shown in Ref.
[10] are bracketed by the results of the old procedure with the values of 
$\sigma  $ parameter equal $0.1 \ GeV$ and $0.14 \ GeV$, we conclude that our new
procedure can describe the data as well. Obviously, the results are meaningful
only for $log_2(Q^2/1GeV^2)>-8$ ($Q^2>0.004GeV^2$). For smaller $Q$ the
statistical fluctuations (visualized by the spread of points) are too large to
draw any conclusions. As already mentioned, the default version of PYTHIA/JETSET
Monte Carlo (without weights) yields a rather flat distribution at the level close to
one. 
\section {Summary and conclusions}
\par
   We have presented a new version of the weight method implementing the BE effect into Monte 
Carlo generators of multiple production, which uses a clustering
procedure for the final state prompt pions.  It provides a better approximation to the (rather 
impractical) full formula (2) than the previous method, although it uses
considerably less computer time. The 
shape of the BE ratio is practically the same for various 
variants of the new method and for the 
old one. This  suggests the stability of the results, and in particular their insensitivity to the 
internal parameter $\epsilon$ of the clustering algorithm. We discuss also how the weights influence 
other distributions. We show that a simple rescaling of
the weights, introduced to restore the shape of the multiplicity 
distribution, reproduces at the same 
time the original inclusive momentum distributions.
\par 
   The satisfactory description of two-particle BE ratio in proton-proton collisions at 630 GeV 
should be obviously just a first step in testing the applicability of our method. In particular, one 
should check if the data for other energies and colliding particles may be described as well 
using the simple form of two-particle weight factor with only one free parameter. One should also 
investigate the three-particle effect and semi-inclusive data (e.g. in bins of restricted multiplicity 
and transverse energy). The BE effect in more than one variable (distinguishing the transverse-, 
longitudinal- and time-like dimensions of the source) may be analyzed by introducing two-
particle weight factor which depends on more variables. Finally, less simplistic distinction of 
direct pions and decay products of various resonances would be welcome.
\par
   It is particularly interesting if the new algorithm may be applied as well to the 
heavy ion collisions (with very high multiplicities $n$) and how the results will compare with the 
description obtained with other methods [7]. Since the computer 
time needed for clustering 
grows only as $n^2$, and the number of clusters is  proportional to $n$, whereas
our previous method [10] required a calculation of $n^5$ terms, we may now expect to deal 
with nuclear data in a reasonable computer time. 
All the applications listed above are under investigation, and for some of them 
encouraging preliminary results are obtained.

\vspace{0.5cm}
{\Large \bf Acknowledgements}
\vspace{0.2cm}
\par
A financial  support from KBN grants No 2 P03B 086 14 and No 2 P03B 196 09
is gratefully acknowledged. 
\vspace{1cm}

{\Large \bf References}
\vspace{0.2cm}         

\par
\noindent 1. D.H. Boal, C.-K. Gelbke and B.K. Jennings, Rev. Mod. Phys. {\bf 62}
(1990) 553, and references therein.
\par
\noindent 2. B. Andersson and W. Hoffman, Phys. Lett. {\bf B169} (1986) 364;
 B. Andersson and \\ M. Ringn\`er,  Nucl. Phys. {\bf B513} (1998) 627.
\par
\noindent 3. T. Sj\"ostrand, Comp. Phys. Comm. {\bf 82} (1997) 1363;
L. L\"onnblad and T. Sj\"ostrand, Phys. Lett. {\bf B351} (1995) 293.
\par
\noindent 4. L. L\"onnblad and T. Sj\"ostrand, Eur. Phys. J. {\bf C2} (1998) 165.
\par
\noindent 5. S. Pratt, Phys. Rev. Lett. {\bf 53} (1984) 1219.
\par
\noindent 6. A. Bia{\l}as and A. Krzywicki,  Phys. Lett. {\bf B 354} (1995) 134.
\par
\noindent 7. Q.H. Zhang et al., e-print nucl-th/9704041.
\par
\noindent  8. S. Jadach and K. Zalewski,  Acta Phys. Pol. {\bf B 28} (1997) 1363.
\par
\noindent 9. V. Kartvelishvili, R. Kvatadze and R. M{\o}ller, reprint
MC-TH-97/04-rev, e-print \\ hep-ph/9704424 v2.
\par
\noindent 10. K. Fia{\l}kowski and R. Wit,  preprint TPJU-3/97, e-print
hep-ph/9703227, Eur. Phys. J. {\bf C}, to be published.
\par
\noindent 11. K. Fia{\l}kowski and R. Wit, Acta Phys. Pol., {\bf B28} (1997)
2039.
\par
\noindent 12. J. Wosiek, Phys. Lett. {\bf B399} (1997) 130. 
\par
\noindent 13. T. Sj\"ostrand and M. Bengtsson, Comp. Phys. Comm. {\bf 43} (1987) 367;
T. Sj\"ostrand, CERN preprint CERN-TH.7112/93 (1993), T. Sj\"ostrand and M. Bengtsson, 
Comp. Phys. Comm. {\bf 46} (1987) 43.
\par
\noindent 14. N. Neumeister et al., Z. Phys.  {\bf C60} (1993) 633.
\par
\noindent 15. J. Pi\u{s}\'{u}t, N. Pi\u{s}\'{u}tov\'{a}, B. Tom\'{a}\u{s}ik,
Phys.Lett. {\bf B368} (1996) 179; Acta Phys. Slov. {\bf 46} (1996) 517. 
\end{document}